\documentclass[a4paper, 12pt]{article}
\usepackage{comment} 
\usepackage{graphicx}
\usepackage{caption}
\usepackage{color,soul}
\usepackage{fullpage} 
\usepackage{natbib}
\usepackage[T1]{fontenc}
\usepackage{upquote}
\usepackage[dvipsnames]{xcolor}
\setcitestyle{authoryear,open={(},close={)}}
\usepackage{hyperref}
\usepackage{amsmath,amsfonts,amssymb}
\usepackage{longtable}
\usepackage{color,soul}
\usepackage[toc,page]{appendix}
\usepackage{booktabs}
\usepackage{nicematrix}
\usepackage[section]{placeins}
\usepackage{multirow}
\usepackage{tikz}
\usetikzlibrary{fit}
\usepackage{colortbl}
\usepackage{import}
\usepackage{float}
\usepackage{setspace}

\DeclareRobustCommand{\hlcyan}[1]{{\sethlcolor{SkyBlue}\hl{#1}}}
\DeclareRobustCommand{\lime}[1]{{\sethlcolor{lime}\hl{#1}}}
\DeclareRobustCommand{\gold}[1]{{\sethlcolor{Goldenrod}\hl{#1}}}
\DeclareRobustCommand{\apricot}[1]{{\sethlcolor{Apricot}\hl{#1}}}
\DeclareRobustCommand{\orch}[1]{{\sethlcolor{Thistle}\hl{#1}}}

\begin{document}
\title{Tag loss models for capture-recapture data without double tagging}
\author{Katarina Skopalova$^1$ \and Wei Zhang$^2$}
\date{$^1$School of Mathematics and Statistics, University of Glasgow, Glasgow, UK\\
$^2$School of Mathematics and Statistics, University of Sydney, Sydney, NSW, Australia}
\maketitle

\begin{abstract}
\noindent
Standard capture-recapture (CR) models assume that marked individuals do not lose their tags during the sampling period. When this assumption is violated, the estimates of model parameters can be biased. Existing tag-loss models generally rely on double-tagging all or a subset of captured individuals, permanent marks, or auxiliary information on tag retention.  However, double-tagging and permanent marking may be impractical, and reliable auxiliary information may be unavailable or difficult to obtain in many situations. In this paper, we develop a novel latent multinomial modelling framework for analysis of CR data with tag loss, without requiring double-tagging or permanent marking. The proposed model framework applies to both closed-population and open-population Cormack-Jolly-Seber (CJS) settings. Through simulation studies, we demonstrate that key model parameters can be reliably estimated in both settings. We illustrate the application of the method using a dataset from a long-term study on elephant seals. The results suggest that robust inference on tag loss and demographic parameters can be achieved without double tagging, thereby avoiding logistical challenges and broadening the framework's applicability to a wider range of studies.
 
\end{abstract}
\textit{Key words:} capture-mark-recapture, closed population, Cormack-Jolly-Seber model, latent multinomial model, single tagging, tag loss
\newpage

\section{Introduction}

Capture-recapture (CR) data are often collected and analysed to estimate key demographic parameters of wildlife populations such as population size, survival, and recruitment. Estimation of these parameters requires individuals to be uniquely identifiable so that individual encounter histories can be constructed for CR models. Individual identification is typically achieved using: (i) physical, non-permanent tags, such as leg bands and neck collars for birds, ear tags for mammals, or PIT tags for reptiles and fish \citep{seber_capture-recapture_2019}; (ii) permanent marks such as brands or tattoos; or (iii) natural markings, i.e., unique physical features such as stripe patterns, scars, or pigmentation \citep{Silvy2020Marking}.\\

A critical assumption of most CR models \citep[e.g.][]{cormack1964estimates, jolly1965explicit, seber1965note, otis1978statistical} is that individuals do not lose their tags during the sampling period. However, in practice tag loss may occur for various reasons, including tag type and placement, individual animal characteristics \citep{seber_capture-recapture_2019}, incorrect tag application, and tag malfunction \citep[e.g.][]{cowen2005capture, tavecchia_modelling_2012}. If not accounted for, tag loss can bias estimates of population size in closed-population models \citep{seber_tag_1981} and affect parameter estimates and their standard errors in open-population models \citep[e.g.][]{arnason_bias_1981,mcdonald_tag_2003}. More specifically, the Jolly-Seber (JS) estimates of survival and birth, as well as the estimates of standard errors for birth, are biased when homogeneous tag loss occurs, i.e.\ when all marked animals have the same probability of tag loss between two sampling occasions \citep{arnason_bias_1981}, and the precision of all estimates is reduced. Under heterogeneous tag loss, the JS estimates of population sizes can be biased, particularly when capture probabilities are low \citep{mcdonald_tag_2003}.\\

Tag loss rates can be estimated through experiments in which animals are held in enclosures or tanks and tag retention is monitored directly over time \citep[see e.g.][]{morgan1993evaluation, belser_ear_2017}. Alternatively, tag loss can be investigated by double-tagging all or a subset of individuals in the population of interest. These studies account for tag loss either by applying ad-hoc corrections to parameter estimates post-analysis \citep[e.g.][]{seber_tag_1981, arnason_bias_1981} or by incorporating tag loss directly into the model. For example, \citet{hyun_accounting_2012} developed a closed-population model for a two-sample design in which tag loss is estimated by combining single- and double-tagged individuals, with double-tagged individuals carrying either two non-permanent tags or a non-permanent tag and a permanent one. \citet{cowen_jollyseber_2006} developed the Jolly-Seber tag-loss model (JSTL) for studies where a proportion of individuals are double-tagged with two non-permanent tags, allowing tag retention rates to vary by release group and time. For the Cormack-Jolly-Seber (CJS) model involving tag loss, \citet{laake_hidden_2014} developed a hidden Markov model for tag loss that incorporates covariates associated with tag retention. \\

However, studies involving double-tagged individuals often make restrictive assumptions about the loss of non-permanent tags, such as assuming no complete tag loss occurs \citep[e.g.][]{tavecchia_modelling_2012} or assuming independent tag loss between the two tags \citep[e.g.][]{cowen_jollyseber_2006, oosthuizen_cohort_2010, hyun_accounting_2012, malcolm-white_complete_2020}. The assumption of independent tag loss may not hold in practice and is difficult to justify without empirical validation (e.g., using permanent marks to identify total tag loss). Studies validating this assumption found evidence of dependence between loss rates of the two tags. For example, in male black bears, the loss rate of a second ear tag depends on whether the first tag has been lost \citep{diefenbach1998modeling, laake_hidden_2014}. A similar pattern was found in southern elephant seals, where individuals had, on average, a 28.1\% greater likelihood of losing a second tag after losing the first one \citep{mcmahon2009tag,schwarz_implications_2012}. Such dependence might arise from factors such as tag placement, differences in application between the two tags, or individual behaviour. When the independent tag loss assumption does not hold, survival estimates are negatively biased \citep{schwarz_implications_2012, laake_hidden_2014}. Pairing a non-permanent tag with a permanent mark avoids these assumptions; however, permanent marks can be costly or difficult to apply, and often require specialised training \citep{seber_capture-recapture_2019}. \\

In addition to requiring double tagging or permanent marks, existing tag loss models often assume that the effect of recycled individuals, i.e., those recaptured after losing both tags, is negligible. For example, the JSTL model \citep{cowen_jollyseber_2006} assumes that recycled individuals are either recognised upon recapture through permanent or natural marks and therefore not tagged again, or that such instances are very limited and have negligible effects on inference. However, \citet{malcolm-white_complete_2020} found that ignoring recycled individuals can lead to biased population size estimates, particularly when capture, survival, and tag loss probabilities are high, and this effect becomes more noticeable for long-term studies.\\

In this paper, we propose a novel method for modelling tag loss, which does not require individuals to be double-tagged or have permanent tags, avoiding the challenges and unrealistic assumptions associated with existing models. The new modelling framework is applicable to both closed-population and CJS problems. The proposed model falls within a general and broad class of latent multinomial models \citep[LMMs,][]{link2010uncovering}. The LMM framework is particularly well suited to situations where individual identification is latent and a single individual can contribute to multiple records in the observed data, but latent identifications cannot be determined from observed data \citep[e.g.][]{bonner2013mark,zhang2022latent}. This is exactly the case for tag loss problems. A single-tagged individual that loses its tag and is subsequently recaptured and re-tagged is essentially a recycled individual. It contributes two encounter histories in the observed data: one containing the partial history before tag loss, and the other one after tag loss. Within the LMM framework, a latent capture history combining the demographic, observational, and tag-loss processes is constructed for each individual, and the latent history is then deterministically linked to one or more observed histories. In this way, the likelihood of the observed data with a complex dependence structure can be specified. Model fitting can be achieved through the efficient saddlepoint approximation method of \cite{zhang2019fast}.\\



\section{Models and Methods}\label{mod}

\subsection{Assumptions and Notation}

We develop tag loss models for two settings: i) closed-population studies under the classic model $M_t$ \citep{otis1978statistical}, and ii) open-population studies under the CJS model. We denote the corresponding tag loss models as CPTL (closed-population tag loss) and CJSTL (CJS tag loss), respectively.\\

The new modelling framework explicitly accounts for tag loss while retaining assumptions similar to those of the corresponding models without tag loss. Similar to model $M_t$, the CPTL model assumes that the population is closed both geographically (i.e. no immigration or emigration) and demographically (i.e. no births or deaths). Capture probabilities vary across sampling occasions but are assumed equal for all individuals within each occasion. The CJSTL model follows the same population dynamics and observational process as the CJS model \citep{cormack1964estimates, jolly1965explicit, seber1965note}, with the likelihood constructed conditional on each individual's initial capture and marking. The number of individuals first captured on each occasion is known under the standard CJS model. However, this number becomes uncertain for all occasions after the first, as individuals that have lost their tags may be recaptured and recorded as new individuals first marked on later occasions. As a result, the numbers of individuals first captured on all occasions except for the first must be estimated rather than treated as known in the CJSTL model. Under both models, we also assume that individuals may lose their tags at most once and that individuals are independent of each other.\\

Suppose that individuals are sampled from the population over $T$ capture occasions, indexed by $t = 1, \ldots, T$. We define the following model parameters for the CPTL model:
\begin{itemize}
    \item $N$ - population size, i.e. the total number of individuals in the population.
    \item $p_{t}$ - the probability that an individual is captured on occasion $t$. We define the vector of capture probabilities across $T$ occasions as $\boldsymbol{p} = (p_{1}, \ldots, p_{T})^{\intercal}$.
\end{itemize}

For the CJSTL model, we define the following parameters: 
\begin{itemize}
    \item $N_t$ - the number of individuals first captured and marked on occasion $t$. Since
        tag loss can only occur after initial marking, $N_1$, the number of individuals first captured and marked on the first occasion, is known. Let $\boldsymbol{N} = (N_{2}, \ldots, N_{T-1})^{\intercal}$ denote the vector of $T-2$ unknown parameters to estimate.
    \item $p_{t}$ - the probability that an individual is captured on occasion $t$, given that it is in the population. We define the vector of capture probabilities across $T$ occasions as $\boldsymbol{p} = (p_{1}, \ldots, p_{T})^{\intercal}$.
    \item $\phi_{t}$ - the probability that an individual survives from occasion $t$ to occasion $t\!-\!1$. We define the vector of all $T-1$ survival probabilities as $\boldsymbol{\phi} = (\phi_{1}, \ldots, \phi_{T-1})^{\intercal}$.
\end{itemize}

For both the CPTL and CJSTL models, we define the tag loss parameter:
\begin{itemize}
    \item $\alpha_t$ - the probability that a marked individual loses its tag between occasions $t$ and $t+1$, given that it retains its tag on occasion $t$. We define the vector of all $T-1$ tag loss probabilities as $\boldsymbol{\alpha} = (\alpha_{1}, \ldots, \alpha_{T-1})^{\intercal}$.
\end{itemize}

\subsection{Observed Data}

We first consider the observed data for a standard CR study with no tag loss. Suppose there are $n$ individuals captured during the study. Let $\boldsymbol{x}_i = (x_{i1}, \ldots, x_{iT})$ denote the capture history of individual $i=1,\ldots,n$, where $x_{it} = 1$ if the individual is captured on occasion $t=1,\ldots,T$ and $x_{it} = 0$ otherwise. For example, $\boldsymbol{x}_i = (1001)$ represents an individual that was captured on the first occasion, not observed on the second and third occasions, and recaptured on the fourth occasion. Each $\boldsymbol{x}_i$ belongs to the set of $K = 2^T - 1$ distinct observable capture histories, excluding the null history. The observed data $\boldsymbol{y} = (y_1, \ldots, y_K)^{\intercal}$ consist of frequencies of all these capture histories, with $y_k$, $k = 1, \ldots, K$ denoting the number of times that the $k$-th capture history is observed.\\

When tag loss occurs, the observed data retain the same format as in standard CR studies. However, the number of observed histories, $n$, does not necessarily equal the number of individuals captured. This is because one individual may generate multiple observed histories. Specifically, a marked individual that loses its tag may be recaptured later and marked again as a new individual.
Suppose an individual loses its tag between occasions $s$ and $s+1$ and is subsequently recaptured and marked as a new individual on occasion $m$, where $s+1 \leq m \leq T$. The individual then produces two separate observed histories:
\begin{itemize}
    \item $\boldsymbol{x}_i^{(1)} = (x_{i1}, \ldots, x_{is}, 0, \ldots, 0),\ \text{where } x_{i1}, \ldots, x_{is} \in \{0,1\} \ \text{for } s < T$; and
    \item $\boldsymbol{x}_i^{(2)} = (0, \ldots, 0, 1, x_{i{m+1}}, \ldots, x_{iT}),\ \text{where } x_{i{m+1}}, \ldots, x_{iT} \in \{0,1\} \ \text{for } m \geq s+1$.
\end{itemize}
For example, suppose we have a study with $T=8$ capture occasions in which some individual $i$ is captured on occasions 1 and 2, loses its tag between occasions 2 and 3, is later recaptured on occasion 6 as a new individual, and is then captured on occasions 7 and 8. This individual would contribute to the observed data with two distinct histories: $\boldsymbol{x}_i^{(1)} = (11000000)$ and $\boldsymbol{x}_i^{(2)} = (00000111)$.\\ 

The capture histories observed under tag loss have a complicated dependence structure, as multiple histories may be generated by the same individual, but their common origin cannot be identified from the observed data. Thus, it becomes challenging to obtain the likelihood for the data. 


\subsection{Latent States}\label{ls}

To account for the uncertainty in individual identification arising from tag loss, we formulate the model in terms of an underlying latent process, that jointly describes capture, survival, and tag-loss dynamics. We introduce a set of latent states to describe an individual's true but unobserved capture events over $T$ sampling occasions, as shown in Table \ref{tab:latent_states}.
\begin{table}
\centering
\caption{Definitions of latent states under the CPTL and CJSTL models.}
\label{tab:latent_states}
\begin{tabular}{cll}
\toprule
Event & Type & Interpretation \\
\midrule
0 & Pre-tag loss  & Not released; CJSTL model only\\
1 & Pre-tag loss  & Not captured \\
2 & Pre-tag loss  & Captured \\
3 & Post-tag loss & Not captured \\
4 & Post-tag loss & Captured \\
5 & Pre- or post-tag loss  & Not survived; CJSTL model only \\
\bottomrule
\end{tabular}
\end{table}
Note that not all possible combinations of the latent states above are valid latent histories. For example, an individual that has entered state 5 cannot return to other states on later occasions. This will be described in more detail through the transition matrices introduced below. \\

Let $J$ denote the number of all valid latent histories, and let $\boldsymbol{\omega}_{j} = (\omega_{j1}, \dots, \omega_{jT})$ denote the $j$-th latent history for $j=1,\ldots,J$, where $\omega_{jt} \in \{0,1,2,3,4,5\}$ under the CJSTL model and $\omega_{jt} \in \{1,2,3,4\}$ under the CPTL model. For example, for a study with $T=8$ capture occasions, the latent history $\boldsymbol{\omega}_{j}=(12124455)$ represents an individual that was first captured on occasion 2, was not observed on occasion 3, was recaptured on occasion 4, lost its tag between occasions 4 and 5, was subsequently captured as a new individual on occasion 5, was observed again on occasion 6, and did not survive from occasion 6 to 7. This latent history produces two observed histories: $(01010000)$ and $(00000100)$, as only states 2 and 4 produce a capture under the tag loss model.\\



\subsection{Transition Matrices}

To formulate the tag-loss models, we derive the probability of each valid latent history in terms of model parameters. Let $\pi_j = P(\boldsymbol{\omega}_{j}\mid \boldsymbol{\theta})$ denote the probability of latent history $\boldsymbol{\omega}_j$, where $\boldsymbol{\theta}=(\boldsymbol{p}, \boldsymbol{\alpha})$ under the CPTL model and $\boldsymbol{\theta}=(\boldsymbol{p}, \boldsymbol{\alpha}, \boldsymbol{\phi})$ under the CJSTL model. We use two transition matrices to define the probabilities of moving between the latent state from occasion $t$ and $t+1$, 
$\boldsymbol{\Gamma}_t^{CP}$ for the CPTL model and $\boldsymbol{\Gamma}_{t}^{CJS}$ for the CJSTL model:

\begin{equation}
\label{eqn:transition_mat_CPTL}
\boldsymbol{\Gamma}_{t}^{CP} = 
\begin{NiceTabular}{c@{\;}ccccc}
& & \Block{1-4}{Occasion $t+1$} \\ 
  &   & 1   & 2   & 3  & 4\\
\Block{4-1}{\rotate Occasion $t$} 
  & 1 & $(1-\alpha_{t})(1-p_{t+1})$ & $(1-\alpha_{t})p_{t+1}$ & $\alpha_{t}(1-p_{t+1})$ & $\alpha_{t}p_{t+1}$\\ 
       & 2 & $(1-\alpha_{t})(1-p_{t+1})$ & $(1-\alpha_{t})p_{t+1}$ & $\alpha_{t}(1-p_{t+1})$ & $\alpha_{t}p_{t+1}$\\ 
       & 3 & 0 & 0 & $1-p_{t+1}$ & $p_{t+1}$\\ 
       & 4 & 0 & 0 & $1-p_{t+1}$ & $p_{t+1}$\\ 
\CodeAfter \SubMatrix[{3-3}{6-6}]
\end{NiceTabular}
\end{equation}

\begin{equation}
\label{eqn:transition_mat_CJSTL}
\boldsymbol{\Gamma}_{t}^{CJS} = 
\begin{NiceTabular}{c@{\;}cccccc}
& & \Block{1-5}{Occasion $t+1$} \\ 
  &   & 1   & 2   & 3  & 4 & 5\\
\Block{4-1}{\rotate Occasion $t$} 
  & 1 & $\phi_t(1-\alpha_{t})(1-p_{t+1})$ & $\phi_t(1-\alpha_{t})p_{t+1}$ & $\phi_t\alpha_{t}(1-p_{t+1})$ & $\phi_t\alpha_{t}p_{t+1}$ & $1-\phi_t$ \\ 
       & 2 & $\phi_t(1-\alpha_{t})(1-p_{t+1})$ & $\phi_t(1-\alpha_{t})p_{t+1}$ & $\phi_t\alpha_{t}(1-p_{t+1})$ & $\phi_t\alpha_{t}p_{t+1}$ & $1-\phi_t$ \\ 
       & 3 & 0 & 0 & $\phi_t(1-p_{t+1})$ & $\phi_tp_{t+1}$ & $1-\phi_t$\\ 
       & 4 & 0 & 0 & $\phi_t(1-p_{t+1})$ & $\phi_tp_{t+1}$ & $1-\phi_t$\\  
       & 5 & 0 & 0 & 0 & 0 & 1
\CodeAfter \SubMatrix[{3-3}{7-7}]
\end{NiceTabular}
\end{equation}
Note that the transition matrices \ref{eqn:transition_mat_CPTL} and \ref{eqn:transition_mat_CJSTL} apply to occasions $f_j \leq t \leq T-1$, where $f_j$ denotes the occasion on which the individual with latent history $\boldsymbol{\omega}_j$ is first captured and marked. The $(r,s)$ entry $\gamma_{rs}^{(t)}$ of $\boldsymbol{\Gamma}_t^{CP}$ and $\boldsymbol{\Gamma}_{t}^{CJS}$ gives the probability of transitioning from state $r$ on occasion $t$ to state $s$ on occasion $t+1$, i.e. $\gamma_{rs}^{(t)}  = P(\omega_{jt+1} = s \mid  \omega_{jt} = r)$, where $r,s \in \{1,2,3,4\}$ for the CPTL model, $r,s \in \{1,2,3,4,5\}$ for the CJSTL model. \\

The probability $\pi_j$ of latent history $\boldsymbol{\omega}_{j} = (\omega_{j1}, \dots, \omega_{jT})$ under the CPTL model is given as 

\begin{equation}
    \pi_j= \prod_{t=1}^{f_j-1}{(1-p_t)}\times p_{f_j} \times \prod_{t=f_j}^{T-1} \gamma_{\,\omega_{jt},\, \omega_{jt+1}}^{(t)},
\end{equation}
where the term $\prod_{t=1}^{f_j-1}(1-p_t)$ represents the probability of not being observed before occasion $f_j$. For example, the probability of latent history $\boldsymbol{\omega}_j=(11212434)$ is
    \begin{align*}  
    P(11212434) 
    &= \left[(1-p_1)(1-p_2)\right] p_3\left[\gamma_{2,1}^{(3)}\gamma_{1,2}^{(4)}\gamma_{2,4}^{(5)}\gamma_{4,3}^{(6)}\gamma_{3,4}^{(7)}\right] \\
    &= \left[(1-p_1)(1-p_2)\right] p_3
   \left[(1-\alpha_{3})(1-p_{4})(1-\alpha_{4})p_{5}\alpha_{5}p_{6}(1-p_{6})p_{7}\right].
    \end{align*}
As the CJSTL model conditions on the first capture of each individual, we only consider capture events after occasion $f_j$. The probability of latent history $\boldsymbol{\omega}_{j}$ under the CJSTL model then becomes
$$\pi_j= \prod_{t=f_j}^{T-1} \gamma_{\,\omega_{jt},\, \omega_{jt+1}}^{(t)}.$$
For example, 
\begin{align*}
P(11212434) 
&=\gamma_{2,1}^{(3)}\gamma_{1,2}^{(4)}\gamma_{2,4}^{(5)}\gamma_{4,3}^{(6)}\gamma_{3,4}^{(7)} \\
&= 
   \phi_3(1-\alpha_{3})(1-p_{4})
   \phi_4(1-\alpha_{4})p_{5}
   \phi_5\alpha_{5}p_{6}
   \phi_6(1-p_{7})
   \phi_7 p_8.
\end{align*}


The latent states defined in Section \ref{ls} produce a total of $4^T$ possible latent histories under the CPTL model and $6^T$ under the CJSTL model. However, the number $J$ of valid latent histories is smaller than these maximum numbers, as the zero probabilities in the transition matrices impose constraints on the order of states in a valid latent history. For example, once an individual enters state 3 or 4 it cannot return to state 1 or 2. This reflects the assumption that tag loss occurs at most once. Under the CJSTL model, if $f_j > 1$ then $\omega_{jt}=0$ for $1 \leq t \leq f_j$, indicating that the individual has not been captured and marked. And once an individual enters state 5 it remains in that state afterwards.




\subsection{Tag Loss Models}

Let $\boldsymbol{z} = (z_1, \ldots, z_J)^{\intercal}$ denote the vector of latent history frequencies for the CPTL model, where $z_j$ is the frequency of latent history $j = 1, \ldots, J$. Under the latent process, each latent history corresponds to exactly one individual so that $\sum_{i=1}^{J}z_j = N$. Assuming independence between individuals, we have 

\begin{equation}
\label{eqn:multinom}
\boldsymbol{z} \sim \text{Multinomial}(N, \boldsymbol{\pi}),
\end{equation}
with $\boldsymbol{\pi} = (\pi_1, \ldots, \pi_J)^{\intercal}$.\\

Under the CJSTL model, let $\boldsymbol{z}_t = (z_{t1}, \ldots, z_{tJ_t})^\intercal, t=1,\ldots,T-1$, where $J_t$ denotes the total number of latent histories with the first capture and marking on occasion $t$, and $z_{tj_t}$ denotes the frequency of latent history $j_t = 1, \ldots, J_t$. It is straightforward that $\sum_{t=1}^{T-1}J_t = J$. Similar to the standard CJS model, the independence between individuals yields that the distribution of the $N_t$ latent histories for individuals first marked on occasion $t$ is multinomial:
\begin{equation}
\label{eqn:eqn1}
\boldsymbol{z}_t \sim \text{Multinomial}(N_t, \boldsymbol{\pi}_t),
\end{equation}
where $\boldsymbol{\pi}_t = (\pi_{t1}, \ldots, \pi_{tJ_t})^{\intercal}$. It follows that 
\begin{equation}
\boldsymbol{z} = (\boldsymbol{z}_1^\intercal, \ldots, \boldsymbol{z}_{T-1}^\intercal)^\intercal
\end{equation}
follows a product multinomial distribution for the CJSTL model.  \\

The vector $\boldsymbol{z}$ is unobservable for both models, so we cannot use (\ref{eqn:multinom}) directly for inference. The observed data vector $\boldsymbol{y}$ is essentially a linear transformation of latent vector $\boldsymbol{z}$. As discussed in Section \ref{ls}, each latent history can generate one or two observed histories, depending on if tag loss occurs. Since both observed and latent histories are known and fixed for a given $T$, their mapping can be explicitly specified. More specifically, we have 
\begin{equation}
\label{lmm}
\boldsymbol{y} = \boldsymbol{A}\boldsymbol{z},
\end{equation}
where $\boldsymbol{A} = (a_{kj})$ is a known $K \times J$ matrix with $a_{kj} = 1$ if latent history $j$ results in observed history $k$, and $a_{kj} = 0$ otherwise. The CPTL model falls with in the class of latent multinomial models \citep[LMMs;][]{link2010uncovering}. The CJSTL model is essentially a latent product-multinomial model, as considered in \cite{zhang2021maximum}. 

\subsection{Parameter Estimation}

The likelihood of the models specified in (\ref{lmm}) can be obtained by summing over all possible latent configurations for $\boldsymbol{z}$ consistent with the observed data $\boldsymbol{y}$:
\begin{equation}
    L(\boldsymbol{\theta} \mid \boldsymbol{y}) = \sum_{\boldsymbol{z}:\, \boldsymbol{y} =\boldsymbol{A} \boldsymbol{z}}
\Pr\!\left(\boldsymbol{z} \mid \boldsymbol{\theta}\right).
\end{equation}
However, it is in general computationally infeasible to evaluate this likelihood directly as it requires enumerating the set of all latent vectors consistent with the observed data, which is enormous in almost all relevant scenarios \citep{link2010uncovering}. Instead, we apply the maximum likelihood approach of \cite{zhang2019fast} based on the saddlepoint approximation (SPA), which shows great efficiency for different types of LMMs and latent product-multinomial models \citep{zhang2021maximum, zhang2022latent,skopalova2025estimating}. The SPA method approximates the probability density or mass distribution of a random variable using its moment generating function (MGF). For the tag loss models considered in this paper, their MGFs can be obtained analytically; see \cite{zhang2019fast} for more details including the implementation. Maximum likelihood estimates are then obtained through numerical optimisation, with standard errors estimated from the inverse of the observed information matrix.

\section{Simulation Study}\label{sims}

We conducted a simulation study to evaluate the performance of the proposed tag-loss models. We explored various settings of parameter values and generated 1,000 datasets for each setting. We used $N=500$ and $T=5$ capture occasions for all simulations. In the closed-population setting, datasets were generated under a $3^2$ experimental design with capture probabilities and tag loss rates set to low (0.2), medium (0.5), or high (0.8) values. We fitted the CPTL model with time-varying capture probabilities and a constant tag loss rate. For convenience, we denote this model by $p_t \alpha_c$, where subscripts `t' and `c' denote time-varying and constant parameters, respectively. In the CJS setting, capture probabilities and tag loss rates were again set to low (0.2), medium (0.5), or high (0.8) values, while survival probabilities were set to either medium (0.5) or high (0.8). We considered three different CJSTL models: $\phi_tp_t\alpha_c$, $\phi_tp_c\alpha_c$, and $\phi_c p_c\alpha_c$. Results for the CPTL model $p_t\alpha_c$ and the CJSTL model $\phi_tp_c\alpha_c$ with $\phi=0.8$ are presented below, while the remaining results are provided in Section A of the Supplementary Material.\\

Across all parameter settings for the CPTL model $p_t \alpha_c$, the estimators of all capture probabilities in $\boldsymbol{p}$ are unbiased. The corresponding 95\% confidence interval (CI) coverages fall within the 2.5-th and 97.5-th percentiles of the binomial distribution, based on true coverage of 95\%. Mean 95\% CI widths for capture probabilities in $\boldsymbol{p}$ range from $0.07$ to $0.13$. Similarly, the estimators of $\boldsymbol{\alpha}$ are also unbiased and produce close-to-nominal CI coverages across almost all settings. One exception is that when the tag loss rate is high and the capture rate is low (i.e. $\alpha=0.8$ and $p=0.2$), the corresponding CI coverage rate ($98\%$) is above the nominal value. Mean 95\% CI widths for $\boldsymbol{\alpha}$ range from $0.05$ ($\alpha=0.2$, $p=0.8$) to $0.28$ ($\alpha=0.5$, $p=0.2$).\\

\begin{figure}
\centerline{\includegraphics[width=1\linewidth]{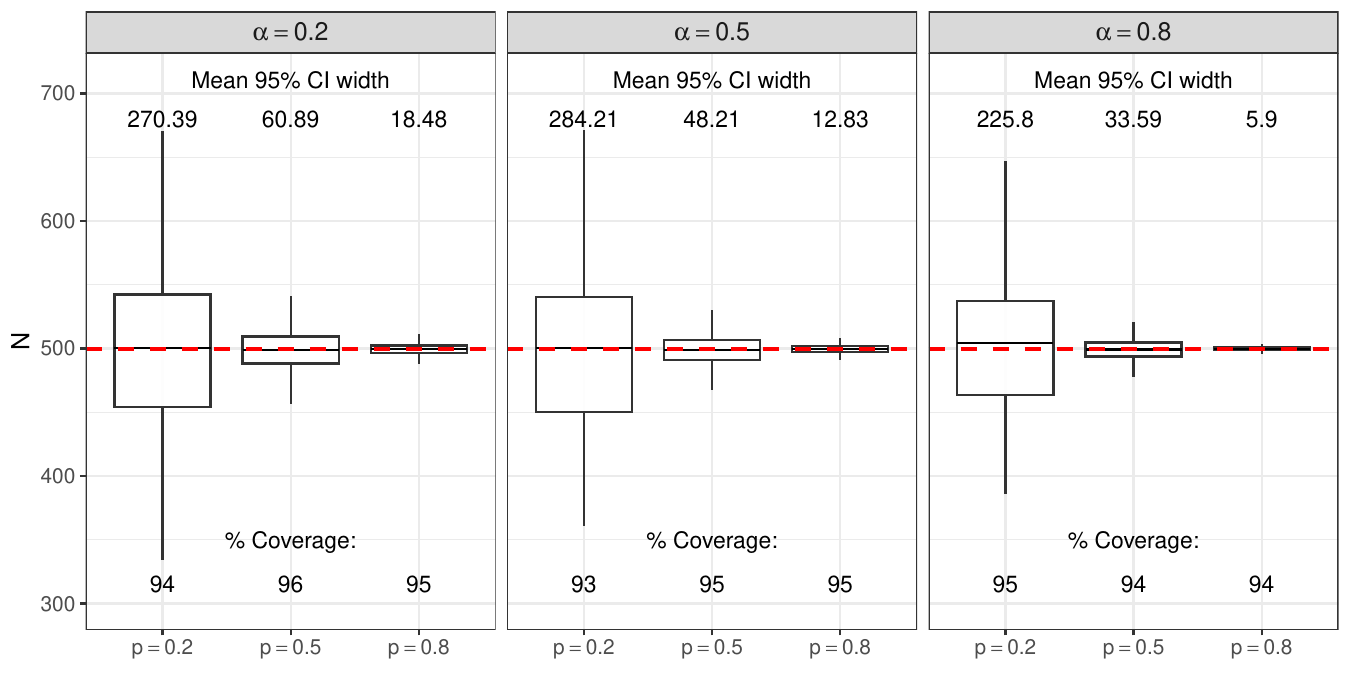}}
    \caption{Boxplots of population size estimates from the CPTL model $p_t\alpha_c$, based on 1,000 simulated datasets under varying tag loss rates $\alpha=0.2, 0.5, 0.8$ and capture probabilities $p=0.2,0.5,0.8$, with $T=5$. The horizontal solid lines within each boxplot represent mean estimates and the red dashed line indicates the true population size $N=500$. Numbers above and below the boxes indicate the mean 95\% CI widths and coverage rates, respectively.}
    \label{fig:plot_n}
\end{figure}

Figure \ref{fig:plot_n} presents the estimation results for population size $N$ under various settings of tag loss rates and capture probabilities for the CPTL model. The estimators of $N$ show negligible bias across all parameter settings. For a given tag loss rate, the precision of population size estimation increases (i.e. mean CI widths decrease) with increasing capture probabilities, which is not surprising as more data are obtained. When capture probabilities are medium or high (i.e. 0.5 or 0.8), the precision of estimation for $N$ also increases as tag loss rate increases. The precision does not show obvious variation across different tag loss rates when capture probabilities are relatively low ($p=0.2$). Across all scenarios, 95\% CI coverage rates generally remain close to the nominal level. Overall, these results suggest that, for a closed-population CR study with tag loss, the CPTL model provides nearly unbiased parameter estimates with good precision across a range of scenarios even without double-tagging.\\

\begin{figure}
\centerline{\includegraphics[width=1\linewidth]{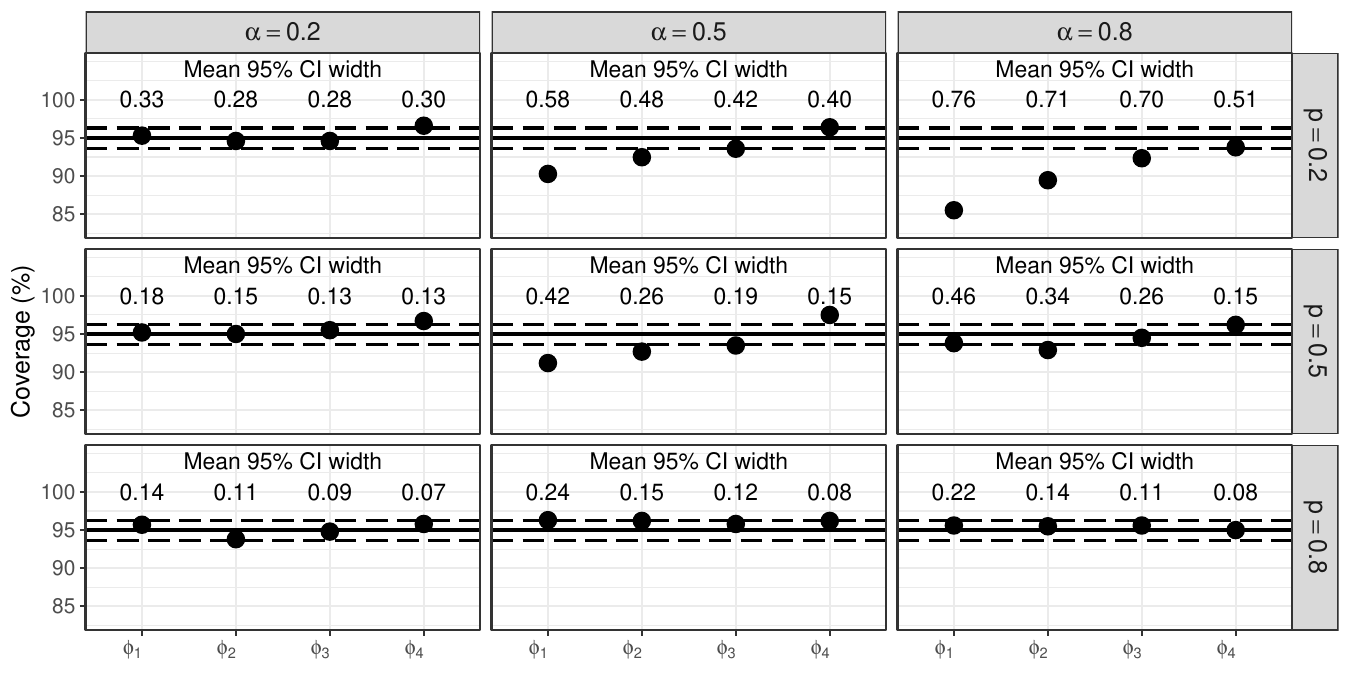}}
    \caption{95\% CI coverage rates and mean 95\% CI widths for survival probabilities from the CJSTL model $\phi_tp_c\alpha_c$  fitted to 1000 datasets under varying tag loss rates ($\alpha=0.2, 0.5, 0.8$) and capture probabilities ($p=0.2,0.5,0.8$), with $\phi=0.8$, $N_t=500$, and $T=5$. The solid horizontal line indicates the nominal coverage level of 95\%, and the dashed lines represent the 2.5-th and 97.5-th percentiles of the binomial distribution for a nominal 95\% coverage level.}
    \label{fig:plot_cjstl}
\end{figure}

We also assessed the performance of the CJSTL model with time-dependent survival probabilities, a constant capture probability, and a constant tag loss rate, i.e. the $\phi_tp_c\alpha_c$ model. Survival probability estimators are unbiased in most scenarios. Minor negative bias was noticed in two settings: i) high tag loss and low capture probability ($\alpha=0.8$, $p=0.2$), where the relative bias was $-6.45\%$ for $\phi_1$, $-2.80\%$ for $\phi_2$, and $2.84\%$ for $\phi_4$; and ii) medium tag loss and low capture probability ($\alpha=0.5$, $p=0.2$), where the relative bias was $-1.64\%$ for $\phi_1$ and $1.53\%$ for $\phi_4$. As shown in Figure \ref{fig:plot_cjstl}, the 95\% CIs exhibit some undercoverage for the survival probability under the two low-capture settings, and tend to be wide. At $\alpha=0.8$ and $p=0.2$, coverage for $\phi_1$ drops by up to $10\%$, while the coverage rates for $\phi_2$ and $\phi_3$ also exhibit some undercoverage. Apart from these two cases, some very minor degree of undercoverage for $\phi_1$ is observed in the setting of $\alpha=0.5$ and $p=0.5$. For the remaining parameter settings, coverage rates generally lie within the theoretical bounds, indicating that survival can be reliably estimated under the CJSTL model in most scenarios.\\

We also checked the estimation results of other model parameters in the two settings discussed above, i.e. $\alpha=0.8, p=0.2$ and $\alpha=0.5, p=0.2$. Estimators of capture probabilities are generally unbiased and achieve nominal coverage across all settings, except that for the setting with $p=0.2, \alpha=0.8$, the 95\% CI coverage rate (92\%) falls slightly below the theoretical lower bound. The tag loss estimators show slight negative bias ($-2.44\%$ and $-1.52\%$) and below-nominal coverage rates ($87\%$ and $92\%$) when $p=0.2,\alpha=0.8$ and $p=0.2,\alpha=0.5$, respectively, but are otherwise unbiased and achieve nominal coverage across the remaining scenarios. The population sizes $N_t$ are generally estimated without bias and achieve the nominal coverage, with minor undercoverage ($93\%$) in settings when $\alpha=0.5, p=0.2$, and $\alpha=0.5, p=0.5$. \\

The simulation results demonstrate that both the CPTL and the CJSTL model exhibit promising performance in estimating parameters of interest, such as survival probabilities and population size. The slight negative bias and undercoverage observed for $\phi_1$ and $\phi_2$ when $\alpha=0.8$ and $p=0.2$ may be attributed to the relatively low capture probability and high tag loss rate considered in this scenario. Under these conditions, individuals released during the first two occasions have a high probability of losing their tags before subsequent recapture occasions, reducing the amount of information available for estimating their survival probabilities. The resulting limited recapture information may lead to less stable parameter estimates and affect the accuracy of the associated CIs, potentially contributing to the observed bias and undercoverage. When tag loss rate is lower or capture probabilities become higher, more recaptures data are available to separate survival from tag loss, making estimation more stable and precise.


\section{Case Study}\label{app}

We apply the CJSTL model to the elephant seal dataset collected as part of a long-term study conducted between 1993 and 2000 on Macquarie Island, Australia \citep{mcmahon2006assessing}. In the study, approximately 6,000 individuals were double-tagged with uniquely numbered plastic flipper tags within 24 hours of birth and subsequently hot-iron branded on their flanks at weaning.\\

Because individuals were double-tagged and branded in the original dataset, we modified the data to apply the single-tag CJSTL model. We assumed individuals could not be recognised by their brands. Individuals captured and recognised based on brands were therefore treated as new individuals assigned with a new tag. The dataset did not contain information on which of the two non-permanent tags was lost, and thus we could not simply select a particular one for the analysis while ignoring the other. Individuals recaptured with only one non-permanent tag were therefore treated as new individuals that were newly tagged.\\

Table \ref{tab:seals_phitpc} presents estimation results of fitting the CJSTL model to the modified seals dataset. We fitted the CJSTL model $\phi_tp_c\alpha_c$, with time-varying survival probabilities, constant capture probability, and constant tag loss rate. The results indicate an estimated tag loss rate of $0.32$ $(95\% \text{ CI=} 0.30, 0.34)$ in this study. Survival probability estimates vary from $0.60$ ($\phi_5$; between 1997 and 1998) to $0.93$ ($\phi_2$; between 1994 and 1995). While these results are not directly comparable to previous analyses of the seals dataset due to the modifications made to the data, they are broadly consistent with previous findings. \citet{mcmahon2009tag} reported that, assuming dependence between tags, the probabilities of reducing from two tags to one or from one to zero in their double-tagging experiment were approximately between $0.20$ and $0.40$, depending on age of the individual. \citet{malcolm-white_complete_2020} reported a constant survival rate of $0.76$ $(95\% \text{ CI=} 0.75, 0.77)$. The capture probability estimate from the CJSTL model ($0.67$; $95\% \text{ CI=} 0.65, 0.68)$ is also consistent with that reported in \citet{malcolm-white_complete_2020} ($0.68$; $95\% \text{ CI=} 0.67, 0.69)$.

\begin{table}
    \centering
    \rowcolors{2}{gray!15}{white}
    \caption{Estimates of survival ($\phi_t$), capture probability ($p$), and tag loss rate ($\alpha$), along with the corresponding 95\% CIs, obtained from fitting the CJSTL model $\phi_tp_c\alpha_c$ to the elephant seal dataset.}
        \begin{tabular}{ll}
        \hline
            Parameter & Estimate (95\% CI) \\
        \hline
            $\phi_1$  & 0.86 (0.80, 0.90)\\
            $\phi_2$  & 0.93 (0.87, 0.97)\\
            $\phi_3$  & 0.67 (0.64, 0.71)\\
            $\phi_4$  & 0.64 (0.60, 0.67)\\
            $\phi_5$  & 0.60 (0.57, 0.63)\\
            $\phi_6$  & 0.63 (0.60, 0.66)\\
            $\phi_7$  & 0.72 (0.69, 0.75)\\
            $p$  & 0.67 (0.65, 0.68)\\
            $\alpha$  & 0.32 (0.30, 0.34) \\
        \hline
        \end{tabular}
    \label{tab:seals_phitpc}
\end{table}

\FloatBarrier
\section{Discussion}\label{dis}

This paper presents a novel method of modelling tag loss in CR data under both closed-population and the open-population CJS settings. Unlike existing approaches, the proposed framework estimates tag loss without requiring double-tagging, permanent marks, or external tag-retention information, and thus avoids the associated practical difficulties or costs and some restrictive model assumptions. In addition, our approach naturally incorporates recycled individuals without using permanent marks. As the model is applicable to standard single-tag datasets, it can be used directly to analyse existing studies and does not require any design modifications in future studies.\\


Simulation studies demonstrated that model parameters can be estimated reliably across a wide range of scenarios under both the CPTL and CJSTL models. We observed reduced performance for the CJSTL model when it combines low capture probabilities, high survival probabilities, and moderate to high tag loss rates. In this scenario, lack of recaptures data makes it challenging to separate survival and tag loss parameters. This is not surprising, as biased inferences also exist in this scenario under the JSTL model even when all individuals are double-tagged \citep{cowen_jollyseber_2006}. This is why \citet{malcolm-white_complete_2020} emphasized that durable tags should be used whenever possible to minimise the probability of tag loss.\\

A limitation of the proposed approach is that it assumes that tag loss is homogeneous, i.e. all animals in the population have the same tag loss rate between two consecutive sampling occasions. Tag loss rates have been shown to depend on time since tagging and may differ between individuals depending on characteristics such as size, age or sex \citep{arnason_bias_1981}. Future work could involve incorporating time since tagging as well as covariates thought to affect tag retention into the models by modelling the logit of the tag loss rate as a linear function of the covariates. \\

Finally, the proposed framework does not account for the possibility that individuals may lose their tags multiple times throughout the study. This may occur when tags have low retention rates, are attached poorly, or for biological reasons related to age, size, or behaviour of the animals. Multiple tag loss can be easily incorporated into the framework by modifying the transition matrices.

\section*{Acknowledgements}
This work was funded by the Carnegie Trust for the Universities of Scotland. We are grateful to Dr. Jafet Belmont Osuna and Dr. Craig Wilkie for their helpful comments on earlier versions of this manuscript.


\begin{thebibliography}{28}
\providecommand{\natexlab}[1]{#1}

\bibitem[{Arnason and Mills(1981)}]{arnason_bias_1981}
A.~N. Arnason and K.~H. Mills. 1981.
\newblock \href {https://doi.org/10.1139/f81-148} {Bias and loss of precision due to tag loss in {Jolly}–{Seber} estimates for mark–recapture experiments}.
\newblock \emph{Canadian Journal of Fisheries and Aquatic Sciences}, 38(9):1077--1095.

\bibitem[{Belser et~al.(2017)Belser, Hewitt, Reitz, Gann, Lewis, and Hellickson}]{belser_ear_2017}
Emily~H. Belser, David~G. Hewitt, Ryan~L. Reitz, Kory~R. Gann, John~S. Lewis, and Mickey~W. Hellickson. 2017.
\newblock \href {https://doi.org/10.1002/wsb.831} {Ear tag and {PIT} tag retention by white-tailed deer}.
\newblock \emph{Wildlife Society Bulletin}, 41(4):752--757.

\bibitem[{Bonner and Holmberg(2013)}]{bonner2013mark}
Simon~J Bonner and Jason Holmberg. 2013.
\newblock Mark-recapture with multiple, non-invasive marks.
\newblock \emph{Biometrics}, 69(3):766--775.

\bibitem[{Cormack(1964)}]{cormack1964estimates}
Richard~M Cormack. 1964.
\newblock Estimates of survival from the sighting of marked animals.
\newblock \emph{Biometrika}, 51(3/4):429--438.

\bibitem[{Cowen and Schwarz(2005)}]{cowen2005capture}
Laura Cowen and Carl~J Schwarz. 2005.
\newblock Capture--recapture studies using radio telemetry with premature radio-tag failure.
\newblock \emph{Biometrics}, 61(3):657--664.

\bibitem[{Cowen and Schwarz(2006)}]{cowen_jollyseber_2006}
Laura Cowen and Carl~J. Schwarz. 2006.
\newblock \href {https://doi.org/10.1111/j.1541-0420.2006.00523.x} {The {Jolly}–{Seber} model with tag loss}.
\newblock \emph{Biometrics}, 62(3):699--705.

\bibitem[{Diefenbach and Alt(1998)}]{diefenbach1998modeling}
Duane~R Diefenbach and Gary~L Alt. 1998.
\newblock Modeling and evaluation of ear tag loss in black bears.
\newblock \emph{The Journal of Wildlife Management}, 62(4):1292--1300.

\bibitem[{Hyun et~al.(2012)Hyun, Reynolds, and Galbreath}]{hyun_accounting_2012}
Saang‐Yoon Hyun, Joel~H. Reynolds, and Peter~F. Galbreath. 2012.
\newblock \href {https://doi.org/10.1080/00028487.2011.639263} {Accounting for tag loss and its uncertainty in a mark–recapture study with a mixture of single and double tags}.
\newblock \emph{Transactions of the American Fisheries Society}, 141(1):11--25.

\bibitem[{Jolly(1965)}]{jolly1965explicit}
George~M Jolly. 1965.
\newblock Explicit estimates from capture-recapture data with both death and immigration-stochastic model.
\newblock \emph{Biometrika}, 52(1/2):225--247.

\bibitem[{Laake et~al.(2014)Laake, Johnson, Diefenbach, and Ternent}]{laake_hidden_2014}
Jeffrey~L. Laake, Devin~S. Johnson, Duane~R. Diefenbach, and Mark~A. Ternent. 2014.
\newblock \href {https://doi.org/10.1007/s13253-014-0190-1} {Hidden {Markov} model for dependent mark loss and survival estimation}.
\newblock \emph{Journal of Agricultural, Biological, and Environmental Statistics}, 19(4):522--538.

\bibitem[{Link et~al.(2010)Link, Yoshizaki, Bailey, and Pollock}]{link2010uncovering}
William~A Link, Jun Yoshizaki, Larissa~L Bailey, and Kenneth~H Pollock. 2010.
\newblock Uncovering a latent multinomial: {A}nalysis of mark--recapture data with misidentification.
\newblock \emph{Biometrics}, 66(1):178--185.

\bibitem[{Malcolm-White et~al.(2020)Malcolm-White, McMahon, and Cowen}]{malcolm-white_complete_2020}
Emily Malcolm-White, Clive~R. McMahon, and Laura L.~E. Cowen. 2020.
\newblock \href {https://doi.org/10.1002/ece3.6052} {Complete tag loss in capture–recapture studies affects abundance estimates: {An} elephant seal case study}.
\newblock \emph{Ecology and Evolution}, 10(5):2377--2384.

\bibitem[{McDonald et~al.(2003)McDonald, Amstrup, and Manly}]{mcdonald_tag_2003}
Trent~L. McDonald, Steven~C. Amstrup, and Bryan F.~J. Manly. 2003.
\newblock \href {https://www.jstor.org/stable/3784604} {Tag loss can bias {Jolly}-{Seber} capture-recapture estimates}.
\newblock \emph{Wildlife Society Bulletin (1973-2006)}, 31(3):814--822.

\bibitem[{McMahon et~al.(2006)McMahon, Burton, Van Den~Hoff, Woods, and Bradshaw}]{mcmahon2006assessing}
Clive~R McMahon, Harry~R Burton, John Van Den~Hoff, Rupert Woods, and Corey~JA Bradshaw. 2006.
\newblock Assessing hot-iron and cryo-branding for permanently marking southern elephant seals.
\newblock \emph{Journal of Wildlife Management}, 70(5):1484--1489.

\bibitem[{McMahon and White(2009)}]{mcmahon2009tag}
Clive~R McMahon and Gary~C White. 2009.
\newblock Tag loss probabilities are not independent: Assessing and quantifying the assumption of independent tag transition probabilities from direct observations.
\newblock \emph{Journal of Experimental Marine Biology and Ecology}, 372(1-2):36--42.

\bibitem[{Morgan and Walsh(1993)}]{morgan1993evaluation}
MJ~Morgan and SJ~Walsh. 1993.
\newblock Evaluation of the retention of external tags by juvenile american plaice (hippoglossoides platessoides) using an aquarium experiment.
\newblock \emph{Fisheries Research}, 16(1):1--7.

\bibitem[{Oosthuizen et~al.(2010)Oosthuizen, De~Bruyn, Bester, and Girondot}]{oosthuizen_cohort_2010}
W.~Chris Oosthuizen, P.~J.~Nico De~Bruyn, Marthán~N. Bester, and Marc Girondot. 2010.
\newblock \href {https://doi.org/10.1111/j.1748-7692.2009.00328.x} {Cohort and tag-site-specific tag-loss rates in mark–recapture studies: {A} southern elephant seal cautionary case}.
\newblock \emph{Marine Mammal Science}, 26(2):350--369.

\bibitem[{Otis et~al.(1978)Otis, Burnham, White, and Anderson}]{otis1978statistical}
David~L Otis, Kenneth~P Burnham, Gary~C White, and David~R Anderson. 1978.
\newblock Statistical inference from capture data on closed animal populations.
\newblock \emph{Wildlife {M}onographs}, (62):3--135.

\bibitem[{Schwarz et~al.(2012)Schwarz, Hindell, McMahon, and Costa}]{schwarz_implications_2012}
L.~K. Schwarz, M.~A. Hindell, C.~R. McMahon, and D.~P. Costa. 2012.
\newblock \href {https://doi.org/10.1890/ES12-00132.1} {The implications of assuming independent tag loss in southern elephant seals}.
\newblock \emph{Ecosphere}, 3(9):art81.

\bibitem[{Seber and Felton(1981)}]{seber_tag_1981}
G.~A.~F. Seber and R.~Felton. 1981.
\newblock \href {https://doi.org/10.1093/biomet/68.1.211} {Tag loss and the {Petersen} mark-recapture experiment}.
\newblock \emph{Biometrika}, 68(1):211--219.

\bibitem[{Seber(1965)}]{seber1965note}
George A~F Seber. 1965.
\newblock A note on the multiple-recapture census.
\newblock \emph{Biometrika}, 52(1/2):249--259.

\bibitem[{Seber and Schofield(2019)}]{seber_capture-recapture_2019}
George A.~F. Seber and Matthew~R. Schofield. 2019.
\newblock \href {https://doi.org/10.1007/978-3-030-18187-1} {\emph{Capture-{Recapture}: {Parameter} {Estimation} for {Open} {Animal} {Populations}}}.
\newblock Statistics for {Biology} and {Health}. Springer International Publishing, Cham.

\bibitem[{Silvy et~al.(2020)Silvy, Lopez, and Peterson}]{Silvy2020Marking}
Nova~J. Silvy, Roel~R. Lopez, and Markus~J. Peterson. 2020.
\newblock Techniques of marking wildlife.
\newblock In Nova~J. Silvy, editor, \emph{The Wildlife Techniques Manual}, 8 edition, volume~1. Johns Hopkins University Press, Baltimore, MD.

\bibitem[{Skopalova et~al.(2025)Skopalova, Osuna, and Zhang}]{skopalova2025estimating}
Katarina Skopalova, Jafet Osuna, and Wei Zhang. 2025.
\newblock Estimating temporary emigration from capture-recapture data in the presence of latent identification.
\newblock \emph{arXiv preprint arXiv:2510.06755}.

\bibitem[{Tavecchia et~al.(2012)Tavecchia, Adrover, Navarro, and Pradel}]{tavecchia_modelling_2012}
Giacomo Tavecchia, Jaume Adrover, Antoni~Muñoz Navarro, and Roger Pradel. 2012.
\newblock \href {https://doi.org/10.1111/j.1365-2664.2011.02074.x} {Modelling mortality causes in longitudinal data in the presence of tag loss: application to raptor poisoning and electrocution}.
\newblock \emph{Journal of Applied Ecology}, 49(1):297--305.

\bibitem[{Zhang et~al.(2019)Zhang, Bravington, and Fewster}]{zhang2019fast}
W~Zhang, M~V Bravington, and R~M Fewster. 2019.
\newblock Fast likelihood-based inference for latent count models using the saddlepoint approximation.
\newblock \emph{Biometrics}, 75(3):723--733.

\bibitem[{Zhang et~al.(2023)Zhang, Bonner, and McCrea}]{zhang2022latent}
Wei Zhang, Simon~J. Bonner, and Rachel~S. McCrea. 2023.
\newblock Latent multinomial models for extended batch-mark data.
\newblock \emph{Biometrics}, 79(3):2732--2742.

\bibitem[{Zhang et~al.(2021)Zhang, Price, and Bonner}]{zhang2021maximum}
Wei Zhang, Steven~J Price, and Simon~J Bonner. 2021.
\newblock Maximum likelihood inference for the band-read error model for capture-recapture data with misidentification.
\newblock \emph{Environmental and Ecological Statistics}, 28(2):405--422.

\end{thebibliography}

\clearpage
\doublespacing      

\renewcommand{\thesubsection}{\Alph{section}.\arabic{subsection}}

\DeclareRobustCommand{\hlcyan}[1]{{\sethlcolor{SkyBlue}\hl{#1}}}
\DeclareRobustCommand{\lime}[1]{{\sethlcolor{lime}\hl{#1}}}
\DeclareRobustCommand{\gold}[1]{{\sethlcolor{Goldenrod}\hl{#1}}}
\DeclareRobustCommand{\apricot}[1]{{\sethlcolor{Apricot}\hl{#1}}}
\DeclareRobustCommand{\orch}[1]{{\sethlcolor{Thistle}\hl{#1}}}



\title{\Large Supplementary Materials for ``Tag loss models for capture-recapture data without double tagging'' by Katarina Skopalova and Wei Zhang.}

\section*{Web Appendix A: Additional simulation results for the CJSTL model}
\addcontentsline{toc}{section}{Web Appendix A: Additional simulation results for the CJSTL model}
\setcounter{table}{0}

\subsection*{Model $\phi_cp_c\alpha_c$}

\begin{figure}[H]
\centerline{\includegraphics[width=1\linewidth]{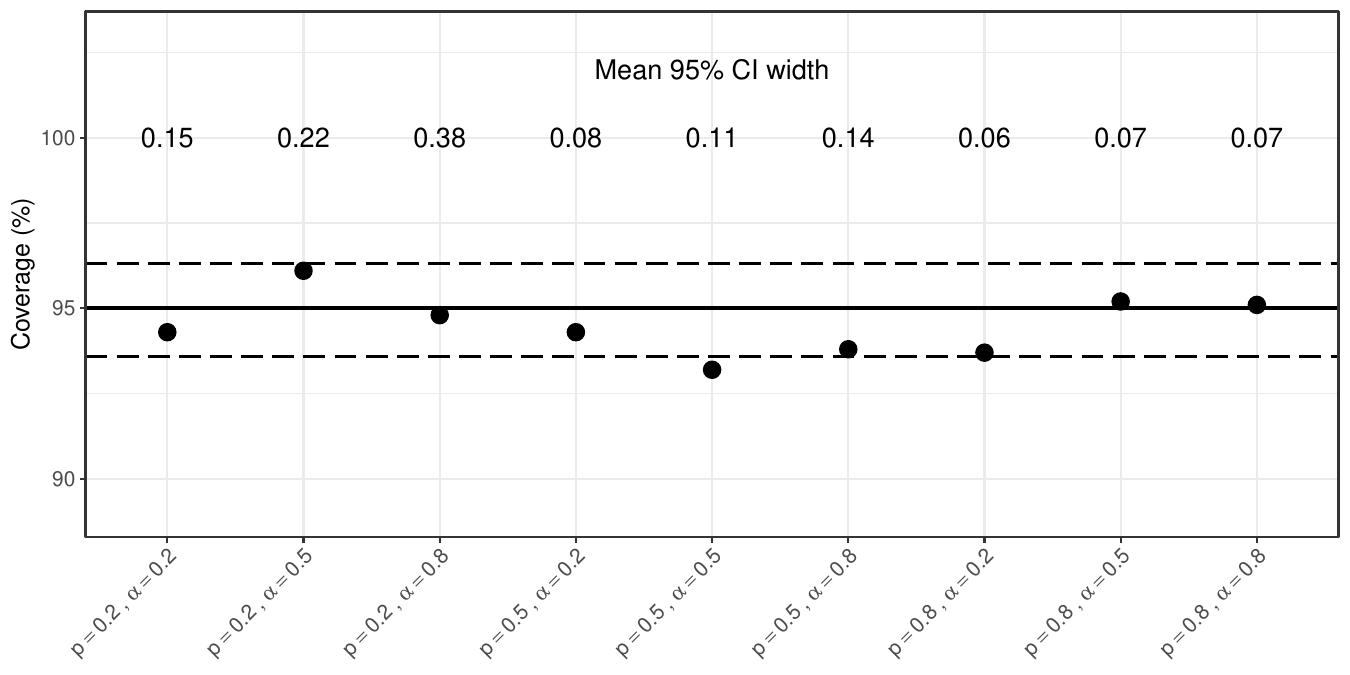}}
    \caption{95\% CI coverage rates and mean 95\% CI widths for survival probabilities from the CJSTL model $\phi_cp_c\alpha_c$  fitted to 1000 datasets under varying tag loss rates ($\alpha=0.2, 0.5, 0.8$) and capture probabilities ($p=0.2,0.5,0.8$), with $\phi=0.6$, $N_t=500$, and $T=5$. The solid horizontal line indicates the nominal coverage level of 95\%, and the dashed lines represent the 2.5-th and 97.5-th percentiles of the binomial distribution based on true coverage of 95\%.}
    \label{fig:CJSTL_phicpc_medphi}
\end{figure} 

\begin{figure}[H]
\centerline{\includegraphics[width=1\linewidth]{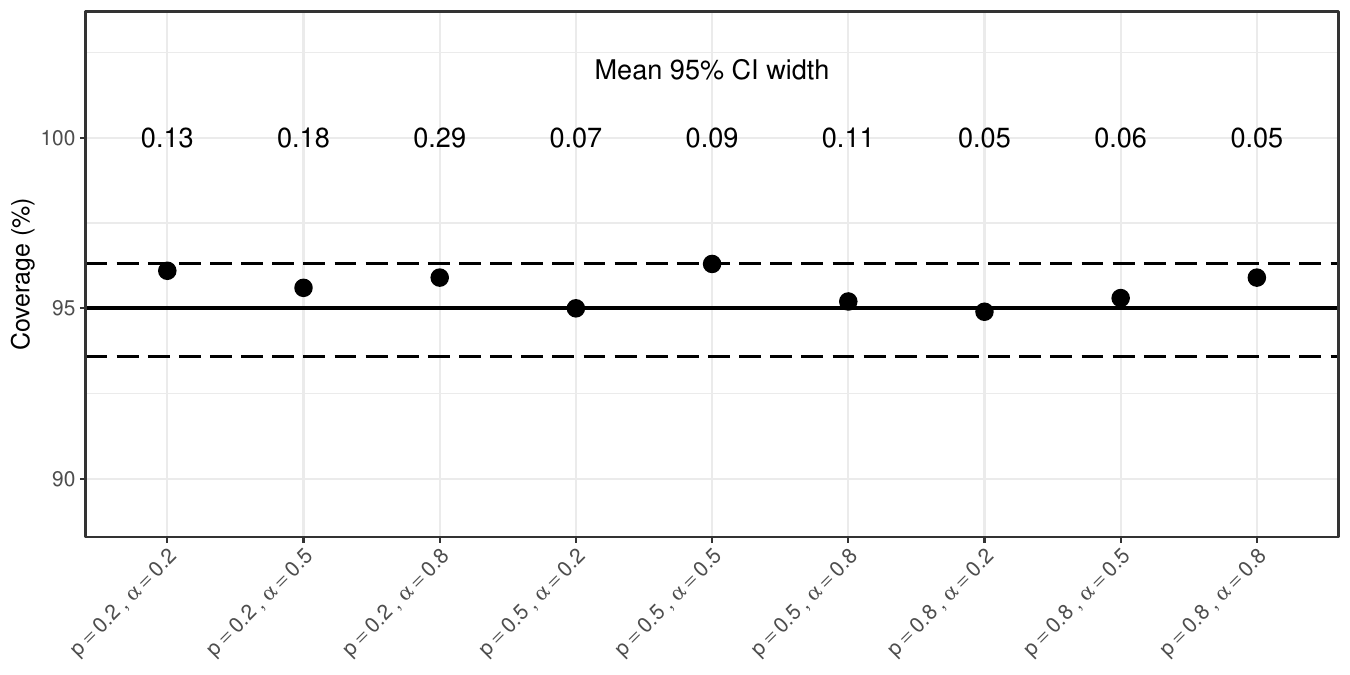}}
    \caption{95\% CI coverage rates and mean 95\% CI widths for survival probabilities from the CJSTL model $\phi_cp_c\alpha_c$  fitted to 1000 datasets under varying tag loss rates ($\alpha=0.2, 0.5, 0.8$) and capture probabilities ($p=0.2,0.5,0.8$), with $\phi=0.8$, $N_t=500$, and $T=5$. The solid horizontal line indicates the nominal coverage level of 95\%, and the dashed lines represent the 2.5-th and 97.5-th percentiles of the binomial distribution based on true coverage of 95\%.}
    \label{fig:CJSTL_phicpc_highphi}
\end{figure} 

\newpage
\subsection*{Model $\phi_tp_c\alpha_c$}

\begin{figure}[H]
\centerline{\includegraphics[width=1\linewidth]{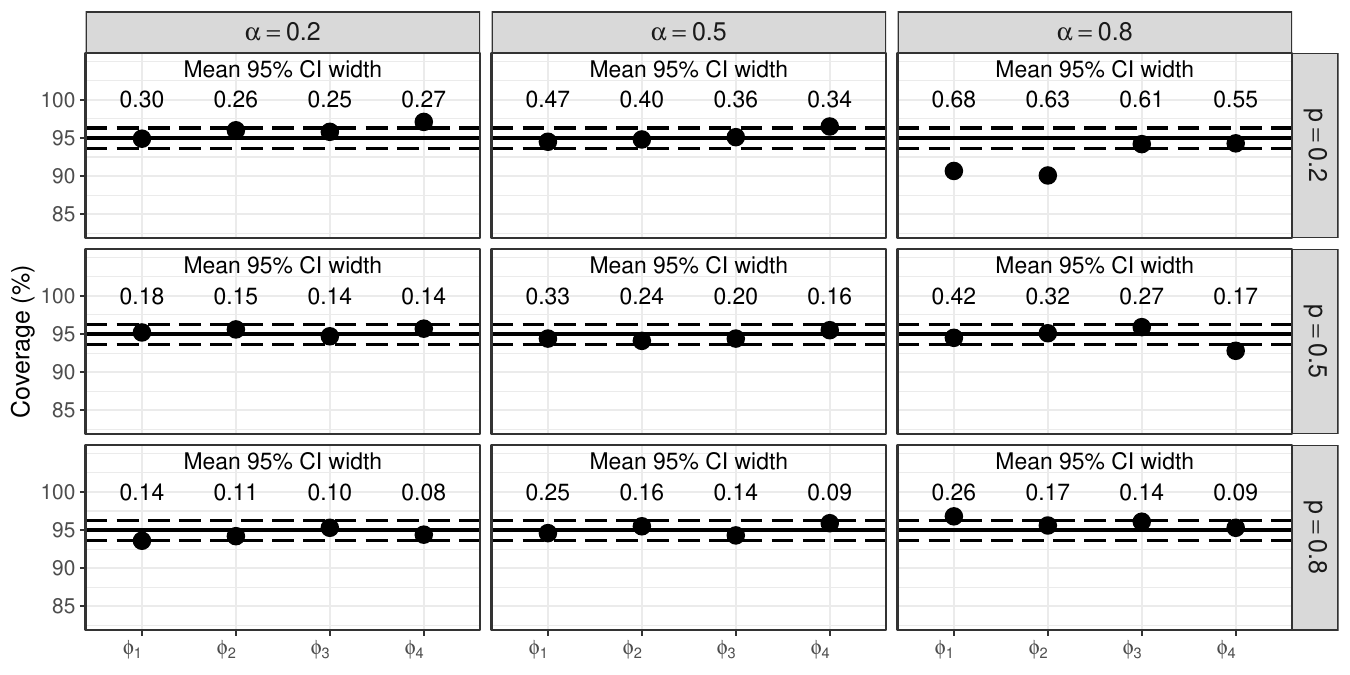}}
    \caption{95\% CI coverage rates and mean 95\% CI widths for survival probabilities from the CJSTL model $\phi_tp_c\alpha_c$  fitted to 1000 datasets under varying tag loss rates ($\alpha=0.2, 0.5, 0.8$) and capture probabilities ($p=0.2,0.5,0.8$), with $\phi=0.6$, $N_t=500$, and $T=5$. The solid horizontal line indicates the nominal coverage level of 95\%, and the dashed lines represent the 2.5-th and 97.5-th percentiles of the binomial distribution based on true coverage of 95\%.}
    \label{fig:CJSTL_phitpc_medphi}
\end{figure} 

\newpage
\subsection*{Model $\phi_tp_t\alpha_c$}

\begin{figure}[H]
\centerline{\includegraphics[width=1\linewidth]{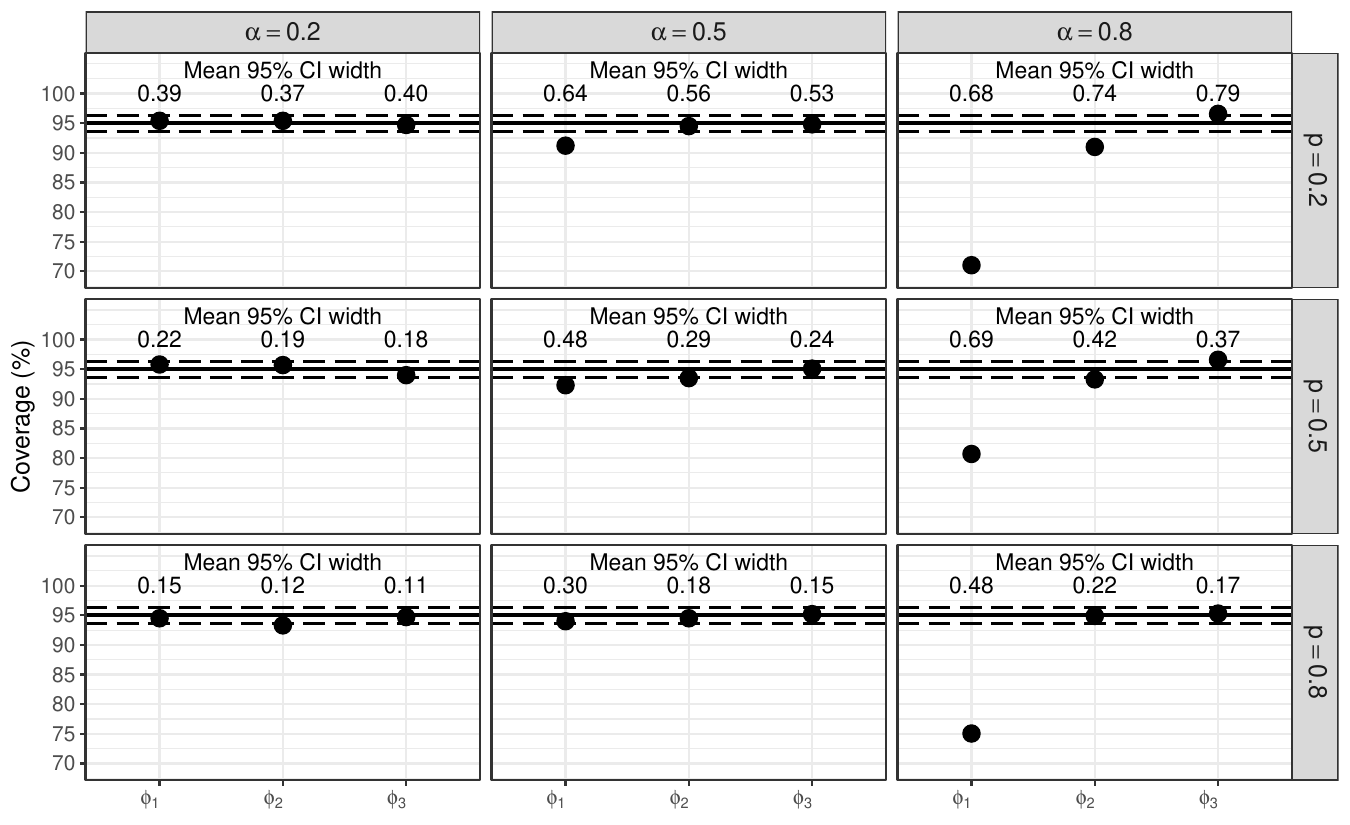}}
    \caption{95\% CI coverage rates and mean 95\% CI widths for survival probabilities from the CJSTL model $\phi_tp_t\alpha_c$  fitted to 1000 datasets under varying tag loss rates ($\alpha=0.2, 0.5, 0.8$) and capture probabilities ($p=0.2,0.5,0.8$), with $\phi=0.8$, $N_t=500$, and $T=5$. The solid horizontal line indicates the nominal coverage level of 95\%, and the dashed lines represent the 2.5-th and 97.5-th percentiles of the binomial distribution based on true coverage of 95\%. Note that $\phi_4$ could not be estimated separately due to confounding with $p_5$ and is therefore not shown.}
    \label{fig:CJSTL_phitpt_prod_medphi}
\end{figure} 

\begin{figure}[H]
\centerline{\includegraphics[width=1\linewidth]{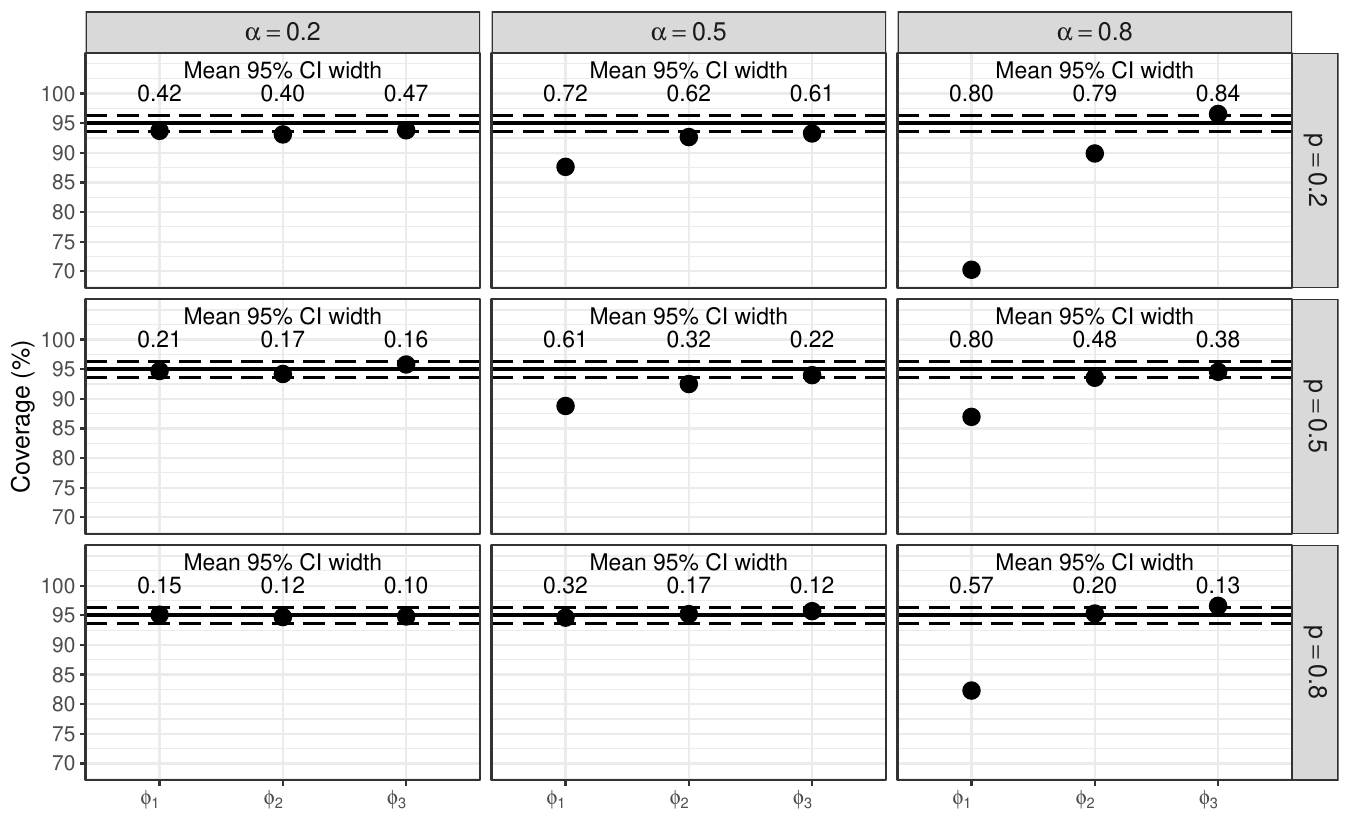}}
    \caption{95\% CI coverage rates and mean 95\% CI widths for survival probabilities from the CJSTL model $\phi_tp_t\alpha_c$  fitted to 1000 datasets under varying tag loss rates ($\alpha=0.2, 0.5, 0.8$) and capture probabilities ($p=0.2,0.5,0.8$), with $\phi=0.8$, $N_t=500$, and $T=5$. The solid horizontal line indicates the nominal coverage level of 95\%, and the dashed lines represent the 2.5-th and 97.5-th percentiles of the binomial distribution based on true coverage of 95\%. Note that $\phi_4$ could not be estimated separately due to confounding with $p_5$ and is therefore not shown.}
    \label{fig:CJSTL_phitpt_prod_highphi}
\end{figure}

\end{document}